\documentclass[twocolumn,amsmath,amssymb,aps,pra,showpacs,tightenlines]{revtex4}
\usepackage{amsmath}
\usepackage{amssymb}
\usepackage{txfonts}
\usepackage{graphicx}% Include figure files
\usepackage{epstopdf}
\usepackage{dcolumn}% Align table columns on decimal point
\usepackage{bm}% bold math
\usepackage[colorlinks=true,citecolor=blue,linkcolor=magenta]{hyperref}
\begin{document}
\title{Phase-mediated magnon chaos-order transition in cavity optomagnonics}
\author{Zeng-Xing Liu}
\author{Cai You}
\author{Bao Wang}
\author{Hao Xiong}
\author{Ying Wu}
\affiliation{School of physics, Huazhong University of Science and Technology, Wuhan 430074, China}
\date{\today}

\begin{abstract}
Magnon as a quantized spin wave has attracted extensive attentions in various fields of physics, such as magnon spintronics, microwave photonics, and cavity quantum electrodynamics. Here, we explore theoretically magnon chaos-order transition in cavity optomagnonics, which is still remains largely unexplored in this emerging field.
We find that the evolution of the magnon experiences the transition from order to period-doubling bifurcation and finally enter chaos by adjusting the microwave driving power.
Different from the normal chaos, the magnon chaos-order transition proposed here is phase-mediated.
Beyond their fundamental scientific significance, our results will contribute to the comprehending of nonlinear phenomena and chaos in optomagnonical system, and may find applications in chaos-based secure communication.
\end{abstract}

\maketitle

%introduction

The strong and even ultrastrong coupling between microwave cavity photons and spin collective excitations has been implemented experimentally in optomagnonical system \cite{4,room0,Photons3,Photons4,Photons5}, in which a millimeter-scale yttrium-iron-garnet (YIG, $\rm{Y_{3}Fe_{5}O_{12}}$) sphere is mounted in a microwave cavity field, and considerable magnon polaritons will generate when the YIG sphere is diametrically pumped \cite{Kerr effect,Bistability}.
YIG is a magnetic insulator well known for its unique material properties \cite{YIG}. On the one hand, YIG has a large spin density as $\rho_{s} \approx 4.22\times10^{27}m^{-3}$ and abundant magnonic nonlinearities. Besides, YIG as a remarkable information carrier can be coupled with microwave photon and acoustic phonon, which provides an outstanding platform for quantum state transfer between different systems \cite{superconducting qubit1,superconducting qubit2,phonon,PT}.
%For example, triple-resonant photon-magnon-phonon coupling has been experimentally acknowledged \cite{phonon}.
On the other hand, YIG maintains a good ferromagnetic property at both cryogenic and room temperatures \cite{room0} as a result of its Curie temperature as high as 559 K.
This novel optomagnonical system derives a new research discipline in the field of magnon spintronics \cite{gradient2}, microwave photonics \cite{Photons3,Photons4,Photons5}, and cavity quantum electrodynamics \cite{4,room0}, as well as leads to many attractive applications in quantum manipulation and quantum information processing \cite{gradient1,APL,computer}.
A vital example is that a broadband and long-lifetime magnon dark state has been observed in a multiple magnon modes coupled a microwave cavity field, which offers a feasible way to establish a magnon gradient memory to store quantum information at room temperature \cite{gradient1}, as well as provides an accessible route to multimode quantum memories and quantum networks.

Lately, the concept of magnon nonlinear effect originating from the magnetocrystalline anisotropy in YIG sphere \cite{anisotropy1,anisotropy2,anisotropy3} has been introduced, and the nonlinearity-induced frequency shift \cite{Kerr effect} and the bistability of cavity magnon polaritons has been observed experimentally \cite{Bistability}.
As a fascinating nonlinear phenomenon, magnon nonlinearity is not only of great significance in studying nonlinear features of the magnetic material, but plays an indispensable role in excavating the potential application of magnonical system as well \cite{sideband1}.
It is well known that nonlinear system is often accompanied by chaotic motion when the nonlinear strength reaches the chaotic threshold \cite{route2}.
A very natural question is whether the magnon nonlinearity can trigger chaos and whether such chaotic behavior can be effectively controlled through a simple as well as feasible way. As far as we know, however, the investigations of the dynamics with a view towards magnon chaotic motion remains largely unexplored, so further insight into the magnon chaotic behavior will contribute to the understanding of nonlinear coherent phenomena and chaos, as well as substantially promote the development of the field of nonlinear magnonics.

In the present work, we explore theoretically magnon chaos in an optomagnonical system, and an efficacious manipulation of the magnon chaos-order transition by bridling the relative phase of the microwave driving field has been discussed in detail.
Here, we mainly focus on the Kittel mode while the other spin wave modes \cite{spin} can be safely neglected, which is demonstrated to be valid in experiment \cite{a}. For one thing, the small YIG sphere can be placed at the uniform field of the antenna to reduce the disturbance from other spin wave modes. For another, the frequency of the Kittel mode can be adjusted to resonate with the microwave cavity field while the other other spin wave modes is off-resonance with the microwave cavity field.
Numerical calculations shown that as we increase the microwave power, the evolution of the magnon polaritons undergoes the route from periodic to period-doubling oscillations and finally enter chaos, which shows an excellent agreement with chaotic theory \cite{route0,route1,route3}.
Compared to normal chaos, the magnon chaos-order transition proposed here is phase-mediated, and we can easily implement the opening and closing of chaotic windows by harnessing the relative phase of the microwave driving field.
%Many interesting physical phenomena are often overlooked when we are doing some perturbation approximations, and thus, the magnon dynamics analysis in this work based on non-perturbative approach can accurately reflect the chaotic characteristics of the magnonical system.
Additionally, the experimental platform shown in Ref. \cite{Bistability} provides all accessible experimental conditions that trigger magnon chaos as well as achieve chaos-order transition. %Furthermore, all these features discussed here can be measured at room temperature due to the Curie temperature of YIG as high as 559 K \cite{room0}.
Our results, therefore, not only deepen our cognition of magnon-polaritons nonlinearity and chaos, but also may provide an accessible platform for achieving chaotic transfer of information processing and chaos-based secure communication, as \cite{0,000,communction1} have suggested.

\begin{figure}[htb]
\centering\includegraphics [width=1\linewidth] {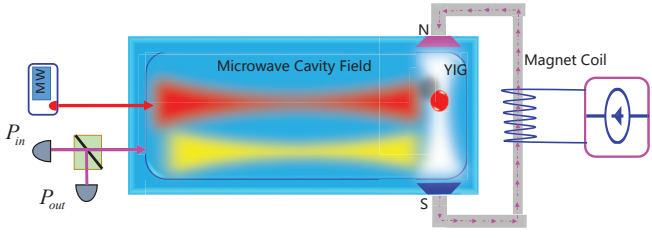}
\caption{Schematic diagram of an optomagnonical system, in which a millimeter-scale YIG sphere is mounted in a high-finesse microwave cavity and placed inside an external bias magnetic field. The YIG sphere is directly driven by a microwave source (MW) and the cavity field is driven by a bichromatic input field.}
\label{fig:1}
\end{figure}

The physical model is schematically shown in Figure \ref{fig:1}.
The system consists of a high-finesse microwave cavity field, in which a millimeter-scale YIG sphere is directly driven by a microwave source with central frequency $\omega_{\rm{d}}$, pump power $P_{\rm{d}}$, initial phase $\varphi_{\rm{d}}$, and amplitude $\xi_{\rm{d}}$ \cite{Kerr effect}. Experiments have demonstrated that considerable  magnon polaritons will generate from the YIG sphere and the magnon nonlinear effect will be exceedingly enhanced by directly pumping the YIG sphere \cite{Kerr effect,Bistability}.
An external bias magnetic field is placed in the microwave cavity field and the frequency of the magnon mode can be tuned at will by adjusting the bias magnetic field \cite{gradient2}.
The interaction between the microwave cavity photon and the magnon mode can be well described by the Hamiltonian as $\hat{H}_{\rm{l}} = \hbar\mathrm{G}(\hat{a}\hat{b}^{\dagger}+\hat{a}^{\dagger}\hat{b})$, where $\mathrm{G}$ is the magnon-photon coupling strength, and $\hat{a}$ ($\hat{a}^{\dagger}$) and $\hat{b}$ ($\hat{b}^{\dagger}$) are the annihilation (creation) operators of the microwave photon with frequency $\omega_{a}$ and the magnon mode with frequency $\omega_{b}$, respectively.
The evolutionary dynamics of the optomagnonical system can be well written by a group of nonlinear partial differential equations as \cite{Kerr effect}:

\begin{eqnarray}\label{equ:0}
   \partial\cdot\pounds &=& \mho\cdot\pounds - i \wp,
\end{eqnarray}
where $\partial\cdot\pounds = (d{\hat{a}}/dt, d{\hat{b}}/dt)^{\mathrm{T}}$, $\pounds = (\hat{a}, \hat{b})^{\mathrm{T}}$, $\wp = (\xi_{\rm{l}}e^{-i(\Delta_{\rm{l}}t+\varphi_{\rm{l}})}+\xi_{\rm{p}}e^{-i(\Delta_{\rm{p}}t+\varphi_{\rm{p}})}, \xi_{\rm{d}}e^{-i\varphi_{\rm{d}}})^{\mathrm{T}}$ with the transpose superscript $\mathrm{T}$, and the coefficient matrix
\begin{eqnarray}
  \mho &=& \left(
  \begin{array}{cc}
   -(i\Delta_{a}+{\kappa}/{2}) & -i\mathrm{G} \\
   -i\mathrm{G} & -(i\Delta_{b}+{\gamma_{b}}/{2})-i\Re \\
  \end{array}
\right), \nonumber
\end{eqnarray}
where $\kappa$ and $\gamma_{b}$, respectively, indicating the decay rate of the microwave cavity field and the magnon mode, are accounted phenomenologically. $\Delta_{\imath (\imath = a, b, \rm{l}, \rm{p})}$ $=$ $\omega_{\imath} - \omega_{\rm{d}}$ are the frequency detuning of the cavity, magnon, and the pump beam modes relative to the microwave driving field, respectively.
$\xi_{\rm{\jmath}}$ $=$ $\sqrt{\kappa P_{\jmath}/\hbar\omega_{\jmath}}$ $(\jmath = \rm{l}, \rm{p})$ are the amplitudes of the bichromatic input field with the frequency $\omega_{\rm{l}}$ and $\omega_{\rm{p}}$, the power $P_{\rm{l}}$ and $P_{\rm{p}}$. $\varphi_{\rm{l}}$ and $\varphi_{\rm{p}}$, respectively, are the initial phases of the dual-pump field, and for convenience, we assume that $\varphi_{\rm{l}}$ = $\varphi_{\rm{p}}$ = $\varphi$.
$\Re = 2\aleph\hat{b}^{\dagger}\hat{b}+\aleph$, where $\aleph=\mu_{0}\aleph_{0}\varrho^{2}/(\mathcal{M}^{2}\mathcal{V}_{m})$ is the magnon nonlinear coefficient with $\mu_{0}$ the magnetic permeability of free space, $\aleph_{0}$ the first-order anisotropy constant,
\begin{figure}[htb]
\centering\includegraphics [width=1\linewidth] {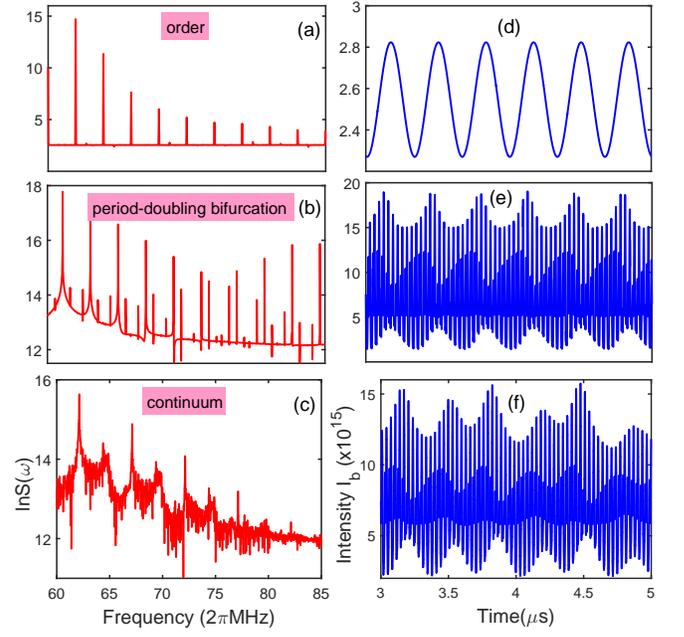}
\caption{ Oscillation of the magnon polaritons is plotted in the frequency domain (a)-(c) as well as in the temporal domain (d)-(f), for three different values of the microwave driving-field power $P_{\rm{d}}$ = 101.3, 184.7, and 229.5 $\rm{mW}$, respectively. %For the sake of simplicity, we only select the frequencies $\omega$ $\in$ (60, 85) $\pi\rm{MHz}$ and the time $t$ $\in$ (3, 5) $\mu s$.
The parameters we used are \cite{Kerr effect}: $\omega_{a}/2\pi$ = $\omega_{b}/2\pi$ = 10.1 $\rm{GHz}$, $\kappa/2\pi$ = 1 $\rm{MHz}$, $\gamma_{b}/2\pi$ = 3 $\rm{MHz}$, $\mathrm{G}/2\pi$ = 44.1 $\rm{MHz}$, $\aleph/\kappa$ $\approx$ $10^{-16}$ \cite{Kerr}. The detuning $\Delta_{i=a, b, \rm{l}, \rm{p}}$ = $\gamma_{b}$, the initial phase $\varphi$ = $\varphi_{\rm{d}}$ = 0, and the powers of the two-tone input field, respectively, $P_{\rm{\rm{l}}}$ = 149.9 $\rm{mW}$ and $P_{\rm{p}}$ = 149.1 $\rm{mW}$. The initial value of $\vec{o}$ = $(a_{r}, a_{i}, b_{r}, b_{i})$ = $(0, 0, 0, 0)$ and $\vec{\delta}=(\delta{a}_{r}, \delta{a}_{i}, \delta{b}_{r}, \delta{b}_{i})$ = $(10^{-8}, 10^{-8}, 10^{-8}, 10^{-8})$.}
\label{fig:2}
\end{figure}
$\varrho$ the gyromagnetic ratio, $\mathcal{M}$ the saturation magnetization, and $\mathcal{V}_{m}$ the volume of the YIG sphere \cite{Kerr effect}.
Equation \ref{equ:0} shows that the nonlinearity of the magnonical system all stems from the magnon Kerr-like nonlinearity, viz., the terms $\aleph\hat{b}^{\dagger}\hat{b}\hat{b}$, and more importantly, the magnon nonlinear coefficient $\aleph$ is inversely proportional to the volume of the YIG sphere \cite{Bistability}, and then the strength of the magnon nonlinearity can be dramatically enhanced by using a small-volume YIG sphere.

In this work, we are only interested in the mean response of the system and thus, the operator can be reduced to their expectation values, viz., $o(t)\equiv\langle\hat{o}(t)\rangle$ \cite{mean}, where $o(t)$ is an any cavity field or magnon operator.
Here, we need to point out that the system ambient temperature we consider is low temperature, so the influence of the thermal noise can be safely ignored.
In order to facilitate the discussion related to chaotic property of the magnon, we define the mean value of each operators as $o=o_{r}+io_{i}$ ($o_{r}$ and $o_{i}$ are real numbers).
The intensity of the magnon polaritons, thence, defined as ${I}_{b}=b_{r}^{2}+b_{i}^{2}$, whose Fourier spectrum $S({\omega})$ in the frequency domain can be acquired by performing the fast Fourier transform on ${I}_{b}$ \cite{sideband2,sideband3}, where $\omega$ is the spectroscopy frequency.

Figure \ref{fig:2} shows the oscillation of the magnon in the frequency domain and temporal domain under different microwave driving fields.
\begin{figure}[htb]
\centering\includegraphics [width=1\linewidth] {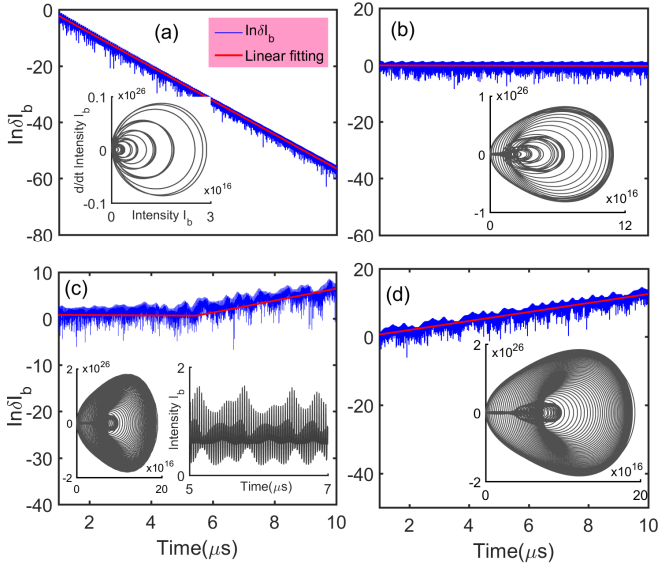}
\caption{ Theoretical calculation of ${\rm{ln}}$$\delta I_{b}$ vary with time ($\mu s$) for four different values of the microwave driving-field power $P_{\rm{d}}$ = 101.3, 184.7, 223, and 229.5 $\rm{mW}$, respectively. The red lines are linear fitting of ${\rm{ln}}$$\delta I_{b}$ and the slope of the fitted line represents the value of the maximal Lyapunov exponent. The insets are the phase-space dynamical trajectories of the magnon obtained by plotting the first derivative of the intensity $I_{b}$ as a function of the intensity $I_{b}$.}
\label{fig:3}
\end{figure}
As shown in Fig. \ref{fig:2}(a), there are higher-order sidebands appear on the frequency spectrogram when the microwave driving laser $P_{\rm{d}}$ = 101.3 $\rm{mW}$.
In this case, the oscillation of the magnon polaritons in the temporal domain is stationary (shown in Fig. \ref{fig:2}(d)), which means that the evolution of the magnon polaritons is regular.
Nonlinear dynamics shows that nonlinear systems often accompanied by chaotic movement when the nonlinear intensity reaches a certain threshold, and period-doubling bifurcation cascades is one of the most common routes to chaos \cite{control}.
Namely, the occurrence of period doubling bifurcation presumably indicates the existence of chaos.
Such period-doubling bifurcation oscillation are obviously observed in Figs. \ref{fig:2}(b) and (e), respectively, when we increase the power of the microwave drive field $P_{d}$ = 184.7 $\rm{mW}$.
Physically, more and more magnon polaritons will be excited when we increase the power of the microwave driving field, which will lead to stronger magnon-photon nonlinear interactions \cite{Kerr effect,Bistability}.
In order to obtain more robust magnon nonlinearity, we further increase the power of the microwave driving field  $P_{d}$ = 229.5 $\rm{mW}$. As Figs. \ref{fig:2}(c) and (f) shown, respectively, the sideband spectra is continuous and the oscillation in temporal domain is aperiodic, which insinuates that the evolution of the magnon is chaotic.
%We note that some non-perturbative signs emerge visibly on the sideband spectrum that is the intensity of the higher-order sidebands have larger than the lower-order sidebands. Our analysis in this work is based on a non-perturbative approach, which may offer a more exhaustive view of nonlinear interaction arising from microwave cavity field and magnon polaritons.

As we all know, the butterfly effect is the most unique feature of chaotic movements, namely, the chaotic system is extremely sensitive to tiny perturbations and the dynamical trajectory in phase space is unpredictable \cite{previous1}.
To investigate quantitatively such hypersensitivity to the initial conditions, we assume the evolution of a perturbation $\vec{\delta}=(\delta{a}_{r}, \delta{a}_{i}, \delta{b}_{r}, \delta{b}_{i})$ which characterizes the divergence degree of adjacent trajectories in phase space.
The maximal Lyapunov exponent (MLE) of the magnon defined by the logarithmic slope of the perturbation $\delta I_{b}$ ($\delta I_{b}=|b+\delta b|^{2}-I_{b}$) versus time $t$ is a mensurable description of the rate of convergence or divergence of nearby trajectories in phase space.

In Fig. \ref{fig:3}(a), a negative MLE indicates that the perturbation $\delta I_{b}$ is attenuate exponentially, consequently, the dynamical trajectory of magnon evolution in phase space with infinitesimally initial perturbation will not diverge but converge to a common fixed point, as shown in illustration.
If, conversely, shown in Fig. \ref{fig:3}(d), the MLE is positive which implies that $\delta I_{b}$ is divergent and the optomagnonical system is extremely sensitive to initial conditions. The initially nearby trajectories in phase space, therefore, will evolve into teetotally different states and becomes greatly complicated and unpredictable, as shown in the inset in Fig. \ref{fig:3}(d).
Furthermore, a zero MLE shown in Fig. \ref{fig:3}(b) demonstrates the appearance of period-doubling bifurcations and the trajectories in phase space as illustration displayed.
In particular, in Fig. \ref{fig:3}(c), we can see that the duration of periodic bifurcation just lasts about 5 microseconds, and after that the oscillation of the magnon becomes chaos, which is referred to as the transient chaos \cite{transit}.
The gray oscillation curve clearly shows the change of the intensity of magnon $I_{b}$ from aperiodic oscillation to periodic oscillation in temporal domain.
From above discussion we can see that as we increase the power of the microwave driving field, the evolution of the magnon polaritons transits from a stabilize behaviours to a periodic-doubling oscillations and finally enter the chaotic state. Physically, magnon chaos-order transition arising from the nonlinear interactions between the microwave photons and magnon polaritons can be substantively modified by the microwave driving source.

Implementing the control of chaos-order transition is of fundamental importance in nonlinear science \cite{control chaos,assist} and may find span-new applications in chaotic encryption and chaos-based secure communication \cite{0,000,communction1}.
In our scheme we find that the magnon chaos-order transition is phase-mediated, namely, one can easily tailor the optomagnonical system enters or withdraws chaotic regime just by harnessing the relative phase of the microwave driving field.
\begin{figure}[htb]
\centering\includegraphics [width=0.95\linewidth] {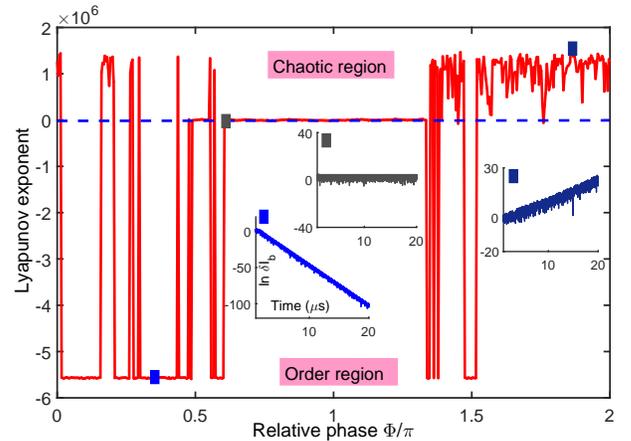}
\caption{The maximal Lyapunov exponent vary with the relative phase of the microwave driving field $\Phi/\pi$ within a fixed time interval $3\rightarrow6$ $\mu s$. The whole area is divided into two region (labeled with the blue dotted line): chaotic region and order region.
The insets: ${\rm{ln}}$$\delta I_{b}$ vary with time ($\mu s$) under different occasions of $\Phi/\pi$ (the blue, gray, and violet dots correspond to the relative phases $\Phi/\pi$ = 0.3567, 0.6052, and 1.8601, respectively). The other parameters are the same as those in Fig. \ref{fig:3}(d)}
\label{fig:5}
\end{figure}
As is well-know that phase is an extremely important physical quantity and plays a critical role in information carrying and coherence characteristics.
The control of chaos-order transition through phase adjustment, therefore, will be very interesting and significant in application.
Figure \ref{fig:5} plots the relationship between the MLE and the relative phase of the microwave driving field. The whole area is divided into two region (labeled with the blue
dotted line): chaotic region and order region. Evidently, the optomagnonical system can be well tailored enters or withdraws chaotic regime by harnessing the relative phase of the microwave driving field.
For example, in the regime of $\Phi/\pi$ $\in$ $(0.3006, 0.4329)$ (blue dot interval), the MLE is negative and ${\rm{ln}}$$\delta I_{b}$ vary with time as the blue line shown, which corresponds to the ordered magnon oscillation. Interestingly, a wide period oscillation window appears in the regime of $\Phi/\pi$ $\in$ $(0.6052, 1.3350)$ (gray dot interval), in which the MLE is zero indicating the period-doubling oscillations of the magnon polaritons.
In addition, a conspicuous chaotic window can be found in the regime of $\Phi/\pi$ $\in$ $(1.5190, 2)$ (violet dot interval), in which the MLE is positive and the value of the MLE characterizing the degree of chaos changes continuously as the alteration of the relative phase $\Phi$.
Excursive oscillation of MLE in the chaotic window fully demonstrates that the evolution of chaotic systems is greatly complicated and unpredictable.

The physical mechanism can be explained as follows: after considering the phase effect, the Hamiltonian of the interaction between the microwave cavity photon and the magnon mode becomes $\widetilde{H}_{\rm{l}} = \hbar(\widetilde{\mathrm{G}}^{\ast}\widetilde{a}\widetilde{b}^{\dagger}+\widetilde{\mathrm{G}}\widetilde{a}^{\dagger}\widetilde{b})$. Here, $\widetilde{a}=\hat{a}e^{-i\varphi}$, $\widetilde{b}=\hat{b}e^{-i\varphi_{\rm{d}}}$, and the magnon-photon coupling strong $\widetilde{\mathrm{G}}=\mathrm{G}e^{-i\Phi}$ with the relative phase of the microwave driving field $\Phi$ = $\varphi-\varphi_{\rm{d}}$. Evidently, the interaction between the microwave photon and the magnon is phase-dependent, namely, the magnon chaos-order transition can be modified by adjusting the relative phase of the microwave driving field.
As Fig. \ref{fig:5} shown, such alternating chaotic and order windows completely determined by the relative phase of the microwave driving field, which provides us an effective and feasible way to manipulate chaos-order transition, and may be important in encryption and decryption of information processing \cite{0,000,communction1}.
%Advantageously, we find that the duration of magnon chaos can also be controlled by manipulating the system cooperativity $\mathcal{C}$. Fig. \ref{fig:5} inset plots the relationship between the duration of magnon chaos and cooperativity $\mathcal{C}/\mathcal{C}_{0}$. Obviously, one can lock the evolution of the magnon to a specific chaotic-temporal domain by adjusting the cooperativity of the system. More importantly, the chaotic time can last as long as 40 $\mu s$ due to the long-lived coherence time of magnon-photon interactions \cite{gradient1}, which may have significance in facilitating the establishment of long-distance chaotic secure communication \cite{Long-distance0,Long-distance1}.

In summary, a fascinating magnon chaotic behavior has been investigated in detail, and an effective method to control chaos-order transition has been discussed.
We found that the dynamic evolution of the magnonical system shows strong parameter dependence, and the magnon chaos-order transition can be well tailored by operating the relative phase of the microwave driving field.
We believe that triggering, transiting, and controlling magnon chaos may have great significance for excavating the nonlinear characteristics of the optomagnonical system. Additionally, such magnon chaos may also be find in other magnonic systems with opto- or electromechanical elements.
Based on the current experimental platform, we believe that the magnon chaos-order transition proposed here will be highly accessible in experiments, so our findings may have the potential to pave the way for exploring important applications in the chaotic encrypt of information and chaos-actuated secure communication.

The work was supported by the National Key Research and Development Program of China (Grant No. 2016YFA0301203) and
the NSF of China (Grant No. 11774113).
\bigskip
\noindent

\end{document}